\documentclass[intlimits,twoside,a4paper]{article}

\usepackage{amsmath,amssymb}
\usepackage{graphicx}
\usepackage{wrapfig}
\usepackage{numprint}
\usepackage[figuresright]{rotating}

\usepackage[T2A]{fontenc}
\usepackage[cp1251]{inputenc}

\usepackage[eqsecnum]{cmpj2}

\issue{2012}{15}{3}{33001}

\doinumber{10.5488/CMP.15.33001}

\title[Asymptotic formulas for integer partitions]
{Asymptotic formulas for integer partitions \\ within the approach of microcanonical ensemble}

\author[D. Prokhorov, A. Rovenchak]{D. Prokhorov, A. Rovenchak}

\address{Department for Theoretical Physics,
Ivan Franko National University of Lviv, \\
12~Drahomanov Str., 79005~Lviv,  Ukraine
}

\authorcopyright{D. Prokhorov, A. Rovenchak, 2012}

\date{Received May 5, 2012, in final form July 9, 2012}

\begin{document}

\maketitle

\begin{abstract}
The problem of integer partitions is addressed using the microcanonical approach which is based on the analogy between this problem in the number theory and the calculation of microstates of a many-boson system. For ordinary (one-dimensional) partitions, the correction to the leading asymptotic is obtained. The estimate for the number of two-dimensional (plane) partitions coincides with known asymptotic results.
\pacs 05.30.Ch, 05.30.Jp
\end{abstract}

\def\be{\begin{eqnarray}}
\def\ee{\end{eqnarray}}
\def\ben{\begin{eqnarray*}}
\def\een{\end{eqnarray*}}

\def\eps{\varepsilon}

\section{Introduction}
In this paper, we address the problem of integer partitions using the physical approach based on  microcanonical treatment.
Partitioning of integers is a problem in the number
theory that originated in the works by Leibniz~\cite{Leibniz1674} and Euler~\cite{Euler1753}. A partition of a positive integer $n$ is a way
of writing $n$ as a sum of positive integers, where the order of
the summands is insignificant. The number of partitions $p(n)$ is
called a {\it partition function}~\cite[Ch.~1]{Andrews76}
Further in this work we refer to $p(n)$ simply as {\it the
number of partitions} to avoid confusion with the respective
physical term.

To clarify the notion of partitions, let us consider the number 5. It can be represented as the following sums:
$$
5 = 4+1
  = 3+2
  = 3+1+1
  = 2+2+1
  = 2+1+1+1
  = 1+1+1+1+1.
$$
Therefore, one has for the number of partitions $p(5) = 7$.

Generalization for higher-dimensional partitions is made in the following fashion. In the $D$-dimensional case, an integer $n$ is
represented as a sum of positive integers $n_{i_1\ldots i_D}$:
\be
n=\sum_{i_1,\ldots,i_D\geqslant0} n_{i_1\ldots i_D}\,,
\ee
where $n_{i_1\ldots i_D}\geqslant  n_{j_1\ldots j_D}$ whenever
$i_1\leqslant j_1$, $i_2\leqslant j_2$, \ldots, $i_D\leqslant j_D$~\cite[p.~179]{Andrews76}.

For instance, in the case of two-dimensional (plane) partitions of 3, one has the following possibilities:
\be\label{PlanePartition3}
\begin{array}{ccrrlr}
3 &\ \ 2\,+\,1 &\ \ 2 &\ \  1\,+\,1\,+\,1 &\ \ 1\,+\,1 &\ \ 1 \\
  &         & + &               &\ \    +  & + \\
  &         & 1 &               &\ \    1  & 1 \\
  &         &   &               &          & + \\
  &         &   &               &          & 1 \\
\end{array}
\ee
yielding the number of plane partitions $p_2(3)=6$~\cite{Andrews76,Almkvist98}.
This can be easily shown by considering the numbers $n_{i_1i_2}$ to be elements of square $1\times1$, $2\times2$, $3\times3$, and so on, matrices. To obtain the sum of matrix elements equal to 3 (provided that they are sorted in a non-ascending order in both rows and columns, according to the above definition), we have:
$$
(3),\quad
\left(\begin{array}{cc}
2&1\\
0&0
\end{array}\right),\quad
\left(\begin{array}{cc}
2&0\\
1&0
\end{array}\right),\quad
\left(\begin{array}{cc}
1&1\\
1&0
\end{array}\right),\quad
\left(\begin{array}{ccc}
1&1&1\\
0&0&0\\
0&0&0
\end{array}\right),\quad
\left(\begin{array}{ccc}
1&0&0\\
1&0&0\\
1&0&0
\end{array}\right).
$$
Note that zero elements are skipped when writing multidimensional partitions.

\section{Physical analogy}\label{secPhysicalAnalogy}
There is a straight analogy between the number of partitions and the number of microstates of a many-boson system of harmonic oscillators.
The approach based on such an analogy was suggested as early as in 1950 by Nanda~\cite{Nanda51}, whose paper is now frequently overlooked, and was later utilized by a number of authors to solve related problems~\cite{Grossman97,Tran04,Comtet07}.

Let the system of one-dimensional harmonic oscillators with spectrum $\eps_j = \hbar\omega j$ have an energy $E$:
\be
E = \hbar\omega\sum_i j_i\,,
\ee
where $j_i$ denotes the quantum number of the $i$th particle.
The summation runs only over the excited states and thus the particles in the ground state (with zero energy) can be arbitrary in number.
The set $\{j_1,j_2,\ldots\}$ corresponds to a certain microstate of the system.
In the quantum case, the particles are indistinguishable, so the permutation of $\{j_1,j_2,\ldots\}$ does not lead to a new microstate. At this point,
we can match the set $\{j_1,j_2,\ldots\}$ and a partition of the number $n=E/\hbar\omega$. Thus, the number of microstates ${\it\Gamma}(E)$ is equal to the number of
partitions of $n$. In the set $\{j_1,j_2,\ldots\}$, any of the numbers $j_i$ can be equal, so one must consider bosonic oscillators.

To make an asymptotic estimation of the number of microstates ${\it\Gamma}(E)$, one can use the well-known Hardy--Ramanujan formula~\cite{Hardy18} for integer partitions:
\be\label{p1d-HR}
p^{\rm HR}(n) = {1 \over 4\sqrt3\,
n}\,\re^{\pi\sqrt{2/3}\,\sqrt n},
\ee
therefore,
\be\label{Gamma}
{\it\Gamma}(E) = {1 \over 4\sqrt3\,
{E/\hbar\omega}}\,\re^{\pi\sqrt{2/3}\,\sqrt{E/\hbar\omega}}.
\ee
In the next section, we will derive this expression from physical considerations and obtain the first correction to this asymptotic result.
For simplicity, we further put the unit of energy $\hbar\omega=1$.

The above considerations might be extended~-- with minor reservations~-- to higher-dimensional partitions, the two-dimensional (plane) ones being best studied. The respective analysis is presented  in section~\ref{SecP2d}.

\section{Corrections to the leading asymptotics of $p(n)$}\label{SecP1d}

Partition function $Z(\beta)$ can be expressed as an integral by introducing the number of microstates ${\it\Gamma}(E)$:
\ben
Z(\beta) = \sum_j \re^{-\beta E_j} =
\int_0^\infty {\it\Gamma}(E)\,
\re^{-\beta E}\, \rd E.
\een
The above expression is nothing but the Laplace transform, so, by inverting it,  we obtain:
\be \label{ZLaplace}
{\it\Gamma}(E) =
{1\over2\pi \ri} \int_{\gamma -\ri\infty}^{\gamma +\ri\infty} Z(\beta)\, \re^{\beta E}\,
\rd\beta.
\ee

The entropy $S(\beta)$ is equal to
\be\label{S=betaE+lnZ}
S(\beta) = \beta E + \ln Z(\beta).
\ee
Thus, we have
\be
Z(\beta) = \re^{S(\beta)}\re^{-\beta E}
\ee
and
\be
{\it\Gamma}(E) ={1\over2\pi \ri} \int_{\gamma-\ri\infty}^{\gamma+\ri\infty} \re^{S(\beta)}\rd\beta.
\ee
Using the Taylor series for entropy in the vicinity of $\beta_0$:
\be\label{S-Taylor}
S(\beta)\simeq S(\beta_0) + {1 \over 2!} S''(\beta_0)(\beta-\beta_0)^2 +{1 \over 3!} S'''(\beta_0)(\beta-\beta_0)^3,
\ee
where $\beta_0$ is the stationary point,
\be
S'(\beta_0)=0,
\ee
for the number of microstates, one obtains:
\be
{\it\Gamma}(E)&\simeq&{\re^{S(\beta_0)}\over2\pi \ri} \int_{\gamma-\ri\infty}^{\gamma+\ri\infty} \exp\left[{{1\over2!}S''(\beta_0)(\beta-\beta_0)^2+{1\over3!}S'''(\beta_0)(\beta-\beta_0)^3}\right]\rd\beta.
\ee
Using the replacement $\beta=\ri x+\beta_0$, we get:
\be
{\it\Gamma}(E)&\simeq&{\re^{S(\beta_0)}\over2\pi } \int_{-\infty}^{\infty}\exp\left[{-{1\over 2!}S''(\beta_0)x^2-{1\over 3!}S'''(\beta_0)\ri x^3}\right]\rd x   \nonumber\\
&=&{\re^{S(\beta_0)}\over2\pi } \int_{-\infty}^{\infty}\re^{-{1\over 2!}S''(\beta_0)x^2}\left\{{\cos\left[{S'''(\beta_0)\over 3!}x^3\right]-\ri \sin\left[{S'''(\beta_0)\over 3!}x^3\right]}\right\}\rd x \nonumber\\
&=&{\re^{S(\beta_0)}\over2\pi } \int_{-\infty}^{\infty}\re^{-{1\over 2!}S''(\beta_0)x^2}\cos\left[{S'''(\beta_0)\over 3!}x^3\right]\rd x.
\ee

This integral can be expressed via the modified Bessel function of the second kind, namely:
\be\label{Gamma-corr}
{\it\Gamma}(E)&\simeq&
{\re^{S(\beta_0)}\over2\pi }
{{2 S''(\beta_0) }\over{\sqrt{3}\left|S'''(\beta_0)\right|}}
\exp\left\{ [S''(\beta_0)]^3\over3[S'''(\beta_0)]^2 \right\}
K_{1/3}\left({ [S''(\beta_0)]^3\over3[S'''(\beta_0)]^2 }\right).
\ee
To calculate the entropy $S(\beta)$, we write the partition function of the oscillator system as follows:
\be
Z(\beta) = \prod_{j=1}^\infty \left(1-\re^{-\beta j}\right)^{-1},
\qquad
\ln Z(\beta) = -\sum_{j=1}^\infty \ln\left(1-\re^{-\beta j}\right).
\ee
Using the Euler--Maclaurin formula to calculate the sum, one obtains~\cite{Tran04}:
\be\label{S-beta-EM}
S(\beta) = \beta E + \ln Z(\beta) = \beta E + \frac{\pi^2}{6\beta} + \frac12\ln\beta -\frac12\ln(2\pi)+\ldots \, .
\ee
Limiting ourselves to the first two terms, the stationary point yields:
\be
\beta_0 = \frac{\pi}{\sqrt{6E}}\,.
\ee
With the same accuracy, we have:
\be
S''(\beta_0) = \frac{2\sqrt6}{\pi} E^{3/2},
\qquad
S'''(\beta_0) = -\frac{36}{\pi^2}E^2.
\ee

Collecting the whole expression (\ref{Gamma-corr}) together, for the number of microstates  we obtain
\be
{\it\Gamma}(E) = \frac{1}{18\sqrt[4]{6}\, E^{3/4}}
\exp\left({28\over 27} \pi \sqrt{2E\over3}\right)
 K_{1/3}\left({1\over 27} \pi \sqrt{2E\over3}\right).
\ee
From this formula, there follows a correction to the main asymptotics of the number of partitions, as one substitutes $E$ with $n$:
\be\label{p1d-corr}
p(n) = \frac{1}{18\sqrt[4]{6}\, n^{3/4}}
\exp\left({28\over 27} \pi \sqrt{2n\over3}\right)
 K_{1/3}\left({1\over 27} \pi \sqrt{2n\over3}\right).
\ee
Taking into account the asymptotic series expansion for large arguments~\cite{AbramowitzStegun}
\be
K_\nu(z) \propto \sqrt{\frac{\pi}2}\,\frac{\re^{-z}}{\sqrt{z}}\left[1+{\cal{O}}\left(\frac1z\right)\right]
\ee
we immediately arrive at the leading asymptotic behavior given by equation~(\ref{p1d-HR}).

\section{Asymptotic behavior of plane partitions}\label{SecP2d}

We consider the energy spectrum of a $2$-dimensional system in the following form:
\be\label{EnergySpectrum}
\eps(j_1,j_2)=j_1+j_2\,.
\ee

To obtain ${\it\Gamma}(E)$, we repeat the derivation presented in the previous section following~\cite{Tran04}, see also~\cite{Rovenchak09}.  Thus, the entropy $S(\beta)$ is equal to
\be
S(\beta) = \beta E + \ln Z(\beta),
\ee
where for energy spectrum (\ref{EnergySpectrum}), the partition function is
\be
Z(\beta) = \prod_{{\rm all\ the\ energies}} \left(1-\re^{-\beta \eps}\right)^{-1} 
=\prod_{j_1=1}^\infty  \prod_{j_2=1}^\infty \left[1-\re^{-\beta (j_1 + j_2)}\right]^{-1}.
\ee
For the logarithm of the partition function $Z(\beta)$, one has:
\be
\ln Z(\beta) = -\sum_{j_1=1}^\infty\sum_{j_2=1}^\infty \ln\left[1-\re^{-\beta (j_1+ j_2)}\right] 
=-\sum_{j=1}^\infty g_j \ln\left[1-\re^{-\beta j}\right],
\ee
where the $j$th level degeneracy is equal to
\be
g_j = j+1 \simeq j,
\ee
therefore,
\be
\ln Z(\beta) =-\sum_{j=1}^\infty j \ln\left[1-\re^{-\beta j}\right].
\ee
After applying the Euler--Maclaurin summation formula, the entropy can be expressed in such a form:
\be
S(\beta) = \beta E + \ln Z(\beta) = \beta E + {\zeta (3) \over \beta^2 } +{1\over 12} \ln \beta
- {1\over  6}\,.
\ee

The standard saddle-point method, for ${\it\Gamma}(E)$ from equation~(\ref{ZLaplace}), yields:
\be
{\it\Gamma}(E) = {\exp[S(\beta_0)] \over \sqrt {2\pi
S''(\beta_0)}}\,.
\ee

As in the previous section, for the stationary point $\beta_0$, so that $S'(\beta_0)=0$, one obtains
\be
\beta_0 = \left[2\zeta(3)\over E\right]^{1/3}
\ee
and
\be
S''(\beta_0) = \frac{3}{[2\zeta(3)]^{1/3}} E^{4/3}.
\ee
Thus, the number of microstates is
\be
{\it\Gamma}(E) = {1 \over  \sqrt {6\pi} }\left[ {2\zeta(3)}\right]^{7/36} E^{-25/36}
\exp\left\{{3\over 2} \left[2\zeta(3)\right]^{1/3} E^{2/3} -{1\over 6}\right\}.
\ee
Substituting energy $E$ with integer $n$, we immediately obtain the result for plane partitions in the following way:
\be\label{p2d-result}
p_2(n) = {1 \over  \sqrt {6\pi} }\left[ {2\zeta(3)}\right]^{7/36} n^{-25/36}
\exp\left\{{3\over 2} \left[2\zeta(3)\right]^{1/3} n^{2/3} -{1\over 6}\right\}.
\ee
Our estimation differs from the result of Wright~\cite{Wright31}
\be\label{p2d-Wright}
p_2^{\rm W}(n) ={1 \over  \sqrt {6\pi} }\left[ {2\zeta(3)}\right]^{7/36} n^{-25/36}
\exp\left\{{3\over 2} \left[2\zeta(3)\right]^{1/3} n^{2/3} + c \right\}
\ee
by a constant factor:
$$
c= \zeta'(-1) = -0.165421\ldots \quad\textrm{versus}\quad -\frac16 = -0.166666\ldots\,.
$$
Note that Wright's asymptotics was later confirmed by Nanda~\cite{Nanda51} with a much more sophisticated analysis of the oscillator system as compared to our approach.

\section{Results and discussion}\label{SecDiscu}

We performed calculations for the number of ordinary partitions $p(n)$ and plane partitions $p_2(n)$ using the expressions (\ref{p1d-corr}) and (\ref{p2d-result}) obtained in this work.

\begin{figure}[!h]
\centerline{%
\includegraphics[angle=-90,scale=0.3]{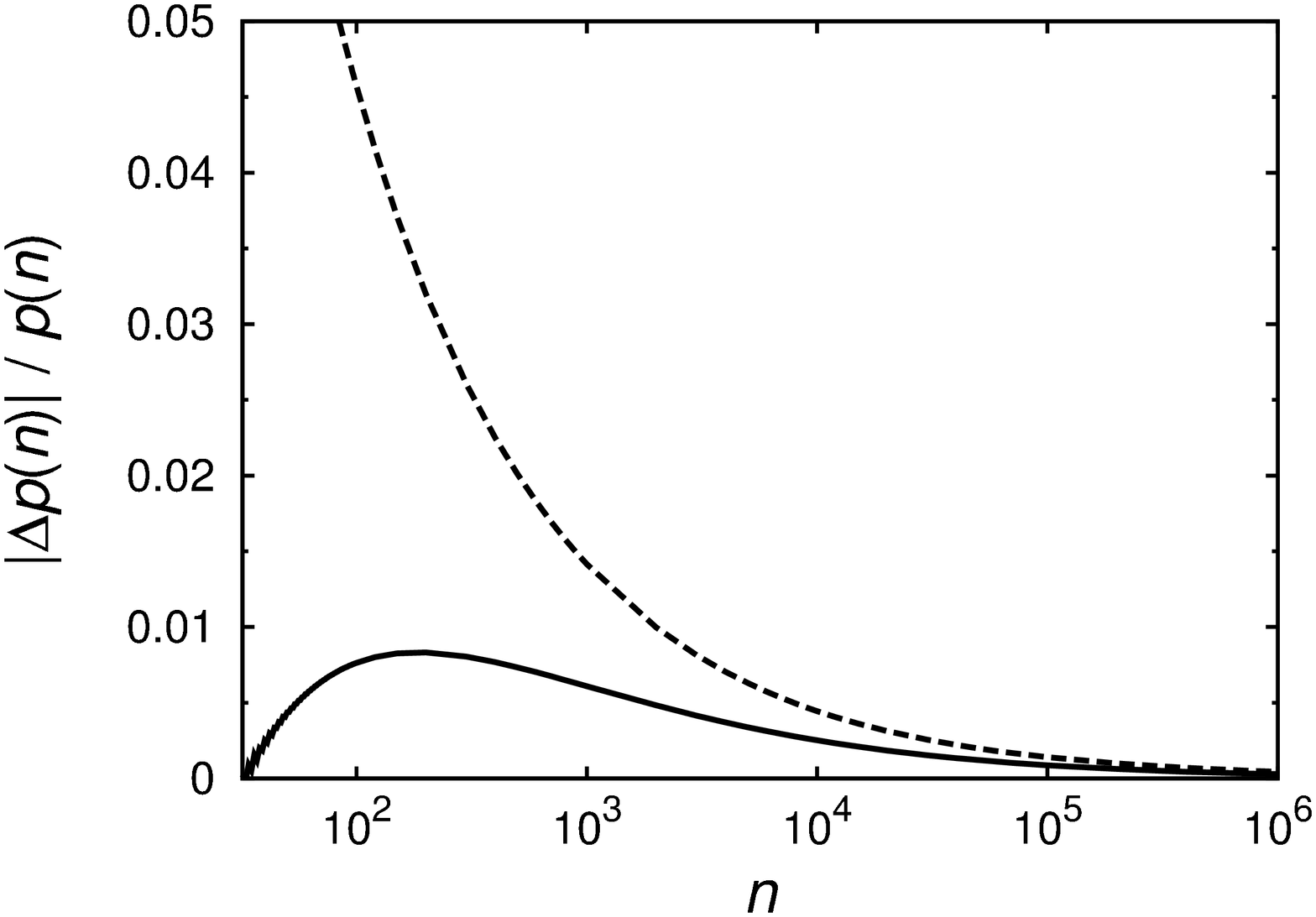}
\
\includegraphics[angle=-90,scale=0.3]{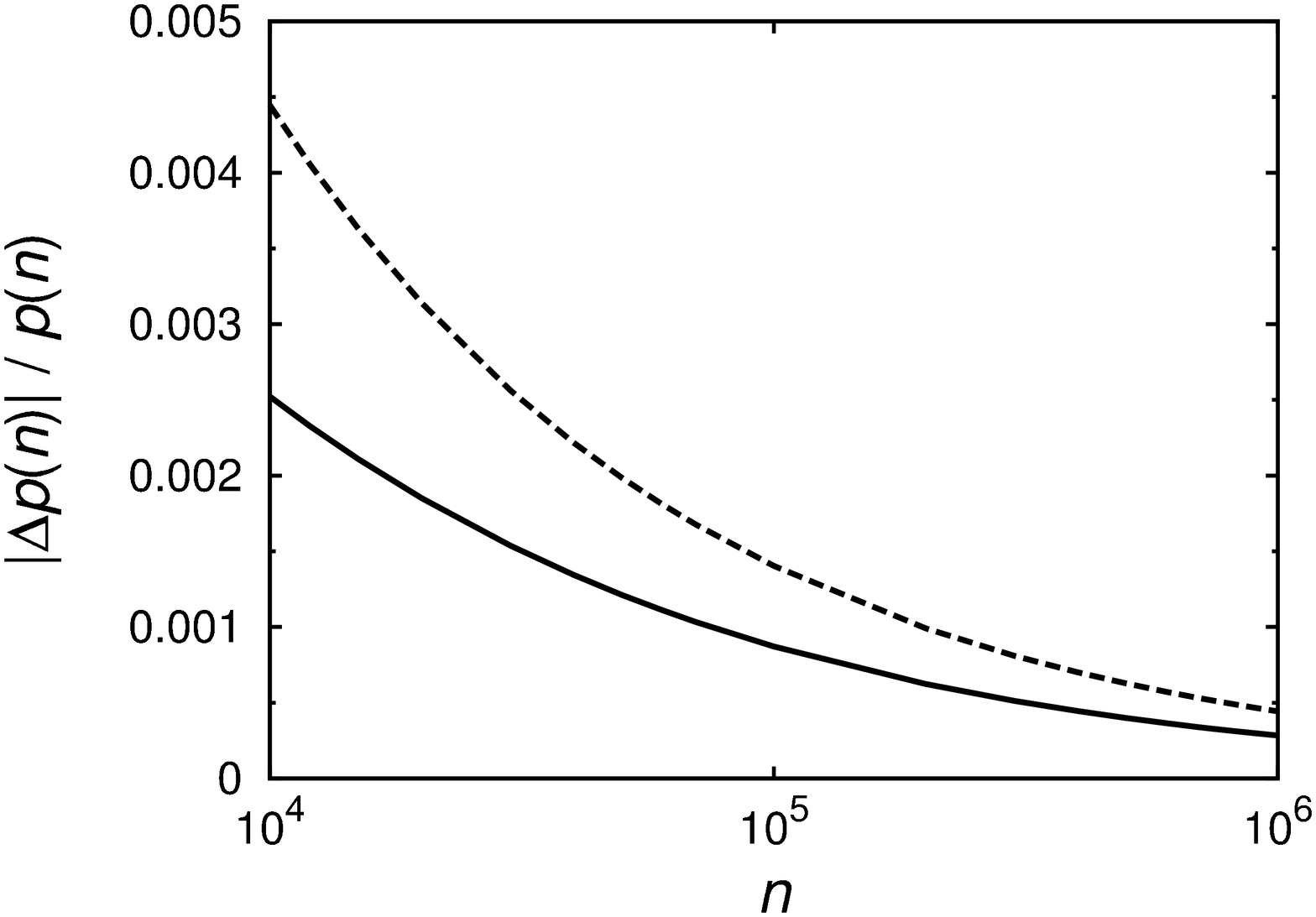}}
\caption{Comparison of relative errors for the estimations of the number of
integer partitions from the real values
(calculated by {\tt wxMaxima 0.8.7}).
Solid curve~-- our result~(\ref{p1d-corr}), dashed curve~-- the leading asymptotics provided by the Hardy--Ramanujan formula (\ref{p1d-HR}). Large $n$ domain in an enlarged view is presented on the right.}
\label{fig:HR-compare}
\end{figure}

The comparison with exact values and the leading asymptotics by Hardy and Ramanujan are given in table~\ref{table:p(n)} and in  figure~\ref{fig:HR-compare}. In the figure, the relative errors are plotted.
For $n=32$ and $n>33$, our correction becomes negative.
For $n>20$, the error is less than one per cent, and it never exceeds seven per cent except for $n=1$, where $p(n)=1$, and our formula yields 2, and $n=5$ with our value of 8 versus $p(n)=5$, which leads to about 15 per cent error. Beyond this domain, the maximum relative deviation of $0.0083285$ is found for $n=186$.

\begin{figure}[!h]
\centerline{%
\includegraphics[angle=-90,scale=0.32]{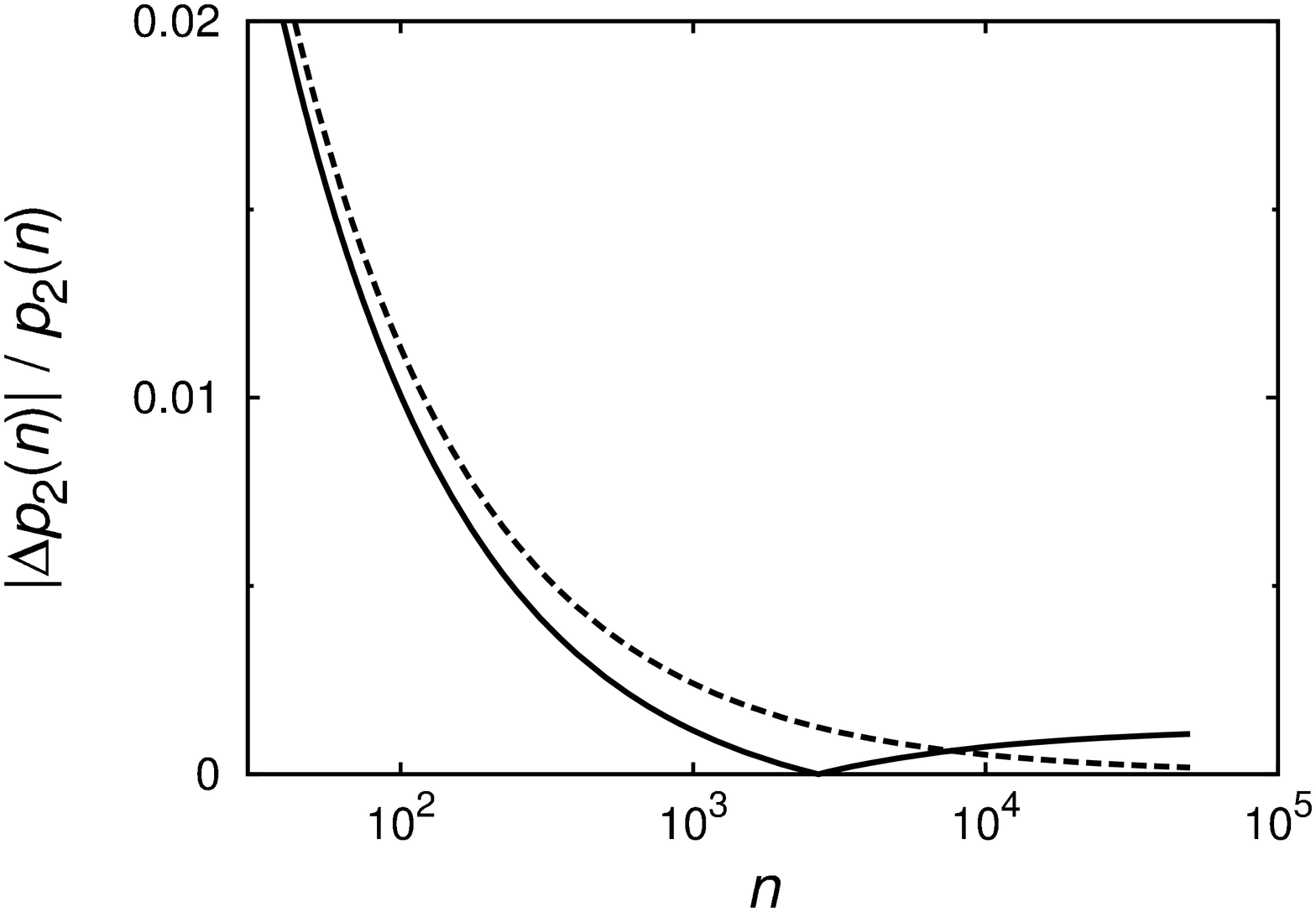}
}
\caption{Comparison of relative errors for the estimations of the number of plane partitions from the real values $p_2(n)$~\cite{Mutafchiev06,PLwww}. Solid curve~-- our result~(\ref{p2d-result}), dashed curve~-- Wright's formula~(\ref{p2d-Wright}).}
\label{fig:PL-compare}
\end{figure}

The plane partitions are demonstrated in table~\ref{table:p2(n)} and in figure~\ref{fig:PL-compare}.
Our result for plane partitions provides a  better estimation compared to that of Wright up to $n=7573$. At $n=2679$, expression~(\ref{p2d-result}) starts to underestimate the number of plane partitions while~(\ref{p2d-Wright}) asymptotically approaches the real values of $p_2(n)$ from above for all $n$.
Interestingly, the approach to the calculation of plain partitions used in this work which takes into account the level degeneracy provides a much better estimate compared to an earlier  suggested~\cite{Rovenchak10} direct treatment of multidimensional oscillators which failed to produce a correct pre-exponential behavior of the power of $n$.

\begin{sidewaystable}
\caption{Number of integer partitions.
Real data $p(n)$ are calculated by {\tt wxMaxima 0.8.7}.
Our result~(\ref{p1d-corr}) is denoted as $p^{\rm our}(n)$.}
\vspace{2ex}
\renewcommand{\arraystretch}{0.9}
\centering
\begin{tabular}{|n{7}{0}|n{32}{0}|n{32}{0}|n{32}{0}|}
\hline
{$\phantom{100}n$}  &   {$\hspace*{2.5cm}\vphantom{\int\limits_0^1}p(n)$}   &
{$\hspace*{2.5cm}p^{\rm HR}(n)$}    &   {$\hspace*{2.5cm}p^{\rm our}(n)$}\\
\hline
1   &   1   &   2   &   2\\
2   &   2   &   3   &   2\\
3   &   3   &   4   &   3\\
4   &   5   &   6   &   5\\
5   &   7   &   9   &   8\\
6   &   11  &   13  &   11\\
7   &   15  &   18  &   16\\
8   &   22  &   26  &   23\\
9   &   30  &   35  &   31\\
10  &   42  &   48  &   43\\
20  &   627 &   692 &   631\\
30  &   5604    &   6080    &   5607\\
40  &   37338   &   40080   &   37244\\
50  &   204226  &   217590  &   203334\\
60  &   966467  &   1024004 &   961084\\
70  &   4087968 &   4312670 &   4061899\\
80  &   15796476    &   16606782    &   15686810\\
90  &   56634173    &   59367760    &   56218131\\
100 &   190569292   &   199280893   &   189113660\\
150 &   40853235313 &   42369336269&    40515857434\\
200 &   3972999029388   &   4100251432188   &   3939941762556\\
300 &   9253082936723602    &   9494094811675004    &   9178612996544433\\
400 &   6727090051741041926 &   6878471626940064454 &   6675403544180355365\\
500 &   2300165032574323995027  &   2346386625611060168255  &   2283287889193750426745\\
600 &   458004788008144308553622    &   466396419561000383265349    &   454786235887791524386094\\
700 &   60378285202834474611028659  &   61401534136286099837866892  &   59970590732043189238732265\\
800 &   5733052172321422504456911979    &   5823869045997298219672106135    &
5695741931526069206199862701\\
900 &   415873681190459054784114365430  &   422080911932431823414681187746  &
413258060326350156090055206821\\
1000    &   24061467864032622473692149727991    &   24401996316802476288263414943062    &
23914862944527589687173572605174\\
10000   &   {$3.61673\times10^{106}$}   &   {$3.63281\times10^{106}$}   &   {$3.60761\times10^{106}$}\\
100000  &   {$2.74935\times10^{346}$}   &   {$2.75321\times10^{346}$}   &   {$2.74695\times10^{346}$}\\
200000  &   {$1.14215\times10^{492}$}   &   {$1.14328\times10^{492}$}   &   {$1.14144\times10^{492}$}\\
500000  &   {$1.52473\times10^{781}$}   &   {$1.52568\times10^{781}$}   &   {$1.52412\times10^{781}$}\\
1000000 &   {$1.47168\times10^{1107}$}  &   {$1.47234\times10^{1107}$}  &   {$1.47127\times10^{1107}$}\\
\hline
\end{tabular}
\label{table:p(n)}
\renewcommand{\arraystretch}{1}
\end{sidewaystable}

\begin{sidewaystable}
\caption{Number of plane integer partitions, real data $p_2(n)$ are presented according to~\cite{Mutafchiev06,PLwww}. Our result is denoted as $p_2^{\rm our}(n)$.}
\vspace{2ex}
\renewcommand{\arraystretch}{0.9}
\centering
\begin{tabular}{|n{5}{0}|n{28}{0}|n{28}{0}|n{28}{0}|}
\hline {$\phantom{10\,}n$} &
{$\hspace*{2.5cm}\vphantom{\int\limits_0^1}p_2(n)$}    &
{$\hspace*{2.5cm}p_2^{\rm W}(n)$} &
{$\hspace*{2.5cm}p_2^{\rm our}(n)$}\\
\hline
1   &   1   & 2  &   2 \\
2   &   3   & 3  &   3 \\
3   &   6   & 7  &   7\\
4   &   13  & 14  &   14\\
5   &   24  & 27  &   27\\
6   &   48  & 51  &   51\\
7   &   86  & 94  &   93\\
8   &   160 & 169  &   169\\
9   &   282 & 300  &   300\\
10  &   500 & 526  &   525\\
15  &   6879 & 7174 &   7165\\
20  &   75278  & 77828   &   77731\\
25  &   696033  & 716466  &   715574\\
30  &   5668963 & 5814929  &   5807691\\
40  &   281846923   & 287805195  &   287446950\\
50  &   10499640707 & 10690144561  &   10676838029\\
60  &   314689799781    & 319730938404  &   319332954078\\
70  &   7937771067795   & 8052251247267  &   8042228226656\\
80  &   173781688194937 & 176070453255058  &   175851289976638\\
90  &   3376508618954817    & 3417560875691699  &   3413306874910366\\
100 &   59206066030052023   & 59876276156314650  &   59801745303809873\\
125 &   52658376905566496345 & 53170908672841062211 & 53104724310582212659\\
150 &   30669139297980503425545 & 30933055614865435615600  &   30894551767372734822993\\
175 &   12797410813969092203145839 & 12896659914533283896907383   & 12880606827741697065333190\\
200 &   4066263490068623016919082185    & 4095085410386804831530514486  &   4089988062550422858471571831\\
1000    &   {$3.54259\times10^{84}$}   & {$3.55112\times10^{84}$} &   {$3.54670\times10^{84}$}\\
10000   &   {$4.50750\times10^{401}$}  & {$4.50983\times10^{401}$}&   {$4.50422\times10^{401}$}\\
100000  &   {$1.11796\times10^{1876}$} & {$1.11808\times10^{1876}$}&   {$1.11669\times10^{1876}$}\\
\hline
\end{tabular}
\label{table:p2(n)}
\renewcommand{\arraystretch}{1}
\end{sidewaystable}

To summarize, the expressions for the number of integer partitions were obtained from the analogy between this number-theoretical problem and the physical problem of calculating the states in a many-boson system within the microcanonical approach.
The correction to the main asymptotics for ordinary (one-dimensional) partitions is shown to give a good estimate even for small numbers.

\section*{Acknowledgements}
We thank Yuri Krynytskyi for discussions on some issues raised in this paper.
The work was partly supported by grant $\Phi\Phi$--$110\Phi$ (No. 0112U001275) from the Ministry of education, science, youth, and sports of Ukraine.

%

\ukrainianpart

\title{Асимптотичні формули для розбиттів цілих чисел у підході мікроканонічного ансамблю}
\author{Д. Прохоров, А. Ровенчак}
\address{Кафедра теоретичної фізики, Львівський національний університет імені Івана Франка, вул.~Драгоманова, 12, Львів, 79005, Україна}

\makeukrtitle

\begin{abstract}
\tolerance=3000%
Розглянуто задачу про розбиття цілих чисел у межах мікроканонічного підходу, який ґрунтується на аналогії між цією задачею з теорії чисел і обчисленням кількості мікростанів багатобозонної системи. Для звичайних (одновимірних) розбиттів отримано поправку до головної асимптотики. Оцінка кількості двовимірних (плоских) розбиттів добре узгоджується з відомими асимптотичними результатами.

\keywords розбиття цілих чисел, плоскі розбиття, бозонні системи

\end{abstract}

\lastpage

\end{document}